\documentclass[12pt]{article}

\usepackage[dvips,final]{graphicx}
\usepackage{epsfig}
\usepackage[center]{subfigure}
\usepackage{tabularx}
\usepackage{hhline}
\usepackage{multirow}
\usepackage{amsmath}

\begin{document}

\begin{flushright}
TTP03-18 \\
SFB/CPP-03-18\\
\end{flushright}
\vspace{0.5cm}
\thispagestyle{empty}
\begin{center}

{\Large \bf
Flavour $SU(3)$ Symmetry 
in Charmless $B$ Decays}
\vskip 1.5true cm

{\large\bf
Alexander~Khodjamirian\,$^{*)}$,
Thomas~Mannel\,, Martin~Melcher\,}
\\[1cm]~\vskip 0.3true cm
{\it Institut f\"ur Theoretische Teilchenphysik, Universit\"at
Karlsruhe,\\  D-76128 Karlsruhe, Germany } \\

\end{center}

\vspace*{1cm}

\begin{abstract}
\noindent
 QCD sum rules are used to estimate
the flavour SU(3)-symmetry violation
in   two-body $B$ decays to pions and kaons.
In  the factorizable amplitudes the    
SU(3)-violation  manifests itself in the ratio of 
the decay constants $f_K/f_\pi$  
and in the differences between the $B\to K$, $B_s\to K$  
and $B \to \pi $ form factors.
These effects are calculated from the QCD two-point 
and light-cone sum rules, respectively, in terms 
of the strange quark mass and the ratio of 
the  strange and nonstrange quark-condensate densities.  
Importantly, QCD sum rules predict that 
SU(3) breaking in the heavy-to-light form factors
can be substantial and 
does not vanish in the heavy-quark mass limit. 
Furthermore, we investigate the strange-quark mass dependence of 
nonfactorizable effects in the  $B\to K\pi$ decay amplitudes. 
Taking into account these effects we estimate the accuracy of    
several SU(3)-symmetry relations between charmless $B$-decay 
amplitudes.
\end{abstract}

\vspace*{\fill}

\noindent $^{*)}${\small \it On leave from
Yerevan Physics Institute, 375036 Yerevan, Armenia} \\

\newpage

\section{Introduction}

The flavour SU(3)-symmetry
is frequently used to reduce and control hadronic uncertainties
in charmless $B$ decays, while analysing various $CP$-related
observables (for a recent comprehensive review see \cite{FleischerPRep}).
The following amplitude relation \cite{GLR} is a well
known example:
\begin{equation}
A(B^-\to \pi^-\bar{K}^0) +\sqrt{2}A(B^-\to \pi^0 K^-)
=\sqrt{2}\left(\frac{V_{us}}{V_{ud}}\right) A(B^-\to \pi^-\pi^0)
\{1+\delta_{SU(3)}\}\,,
\label{rel1}
\end{equation}
where we neglect electroweak penguin contributions
and introduce a parameter
$\delta_{SU(3)}$ to quantify
the SU(3)-violation, so that $\delta_{SU(3)}=0$ in the
exact symmetry limit.

For a reliable use of Eq.~(\ref{rel1}) it is desirable to have a
QCD-based estimate of $\delta_{SU(3)}$.
A usual phenomenological remedy is to relate SU(3) violation
to the ratio of the kaon and pion decay constants $f_K/f_\pi$
and/or to the ratio of $B\to K $ and $B\to \pi$ form factors.
Such estimates, however, rely on the factorization
approximation with its limited
accuracy. Adding nonfactorizable effects,
e.g., in the spirit of QCD factorization \cite{BBNS},
one has the following schematic expression
for a given $B\to P_1P_2$ amplitude
($B=B_{u,d,s}$; $P_{1,\,2}=\pi,K$):
\begin{equation}
A(B\to P_1P_2)
=A_{fact}(B\to P_1P_2)\Bigg\{1+
\frac{\alpha_s C_F}{\pi}\!\!\sum\limits_{i=E,P,A,..} \!\!\delta^{(BP_1P_2)}_i
+\sum\limits_{i=E,P,A,..} \!\!\frac{\lambda_i^{(B P_1P_2)}}{m_B}\Bigg\}\,,
\label{schem}
\end{equation}
where
\begin{equation}
A_{fact}(B\to P_1P_2)
= i\frac{G_F}{\sqrt{2}}(m_B^2-m_{P_1}^2)f_{P_2}f^0_{BP_1}(m_{P_2}^2)
\label{Afact}
\end{equation}
is the factorizable amplitude, $P_2$ being the ``emitted'' meson
with the decay constant $f_{P_2}$, and $f^0_{BP_1}$ is the
$B\to P_1$ transition form factor.
For simplicity, all CKM and short-distance factors
are not shown. The nonfactorizable corrections are
suppressed either by $\alpha_s$ or by inverse powers
of the $b$-quark mass. In Eq.~(\ref{schem}) they are
parametrized by the process-dependent parameters $\delta_i^{(BP_1P_2)}$ and
$\lambda_i^{(BP_1P_2)}$, respectively. The sums indicate
that nonfactorizable contributions  stem from different
effective operators and topologies (emission, penguin, annihilation, etc.).
Moreover, certain decay channels 
receive two factorizable contributions,
so that the term $f_{P_1}f_{B P_2}(m_{P_2}^2)$,
with its nonfactorizable corrections,
has to be added to Eq.~(\ref{schem}).
There are several sources of SU(3) violation
in the $A(B \to P_1P_2)$ amplitudes. The
inequalities $f_K\neq f_\pi$ and $f_{BK}\neq f_{B\pi}\neq f_{B_s K}$
reflect flavour-symmetry breaking
in the factorizable amplitudes.
In addition, differences between the nonfactorizable contributions
may also play a role.  All separate SU(3)-violating
effects have to be accounted and added up
in order to obtain an estimate of $\delta_{SU(3)}$ in Eq.~(\ref{rel1}).

Only  the ratio $f_K/f_\pi$ is known from experiment,
revealing quite a noticeable SU(3) violation:
$f_K=160$ MeV and $f_\pi=131$ MeV.
For heavy-to-light form factors and nonfactorizable
effects one has to rely on theoretical predictions.
Important questions concern therefore
the parametrical dependence of various SU(3)-violation effects
on the quark-mass difference $m_s-m_{u,d}$. We will take into
account all effects
of the first order in $m_s-m_{u,d}$ and in several cases also
those of $O(m_s^2)$ . It is also important
to distinguish the SU(3)-violation effects
proportional to $(m_s-m_d)/m_b$ from those effects
which survive in the $m_b\to\infty$ limit being
of $O((m_s-m_{u,d})/M)$, where $M$
is a large scale independent of the heavy quark mass.

In this paper we investigate the flavour SU(3)- symmetry violation
in charmless $B\to P_1 P_2$ decays in the framework of
QCD sum rules. Within this method, the 
ratios of hadronic matrix elements are calculated 
in terms of  
the quark mass difference $m_s -m_{u,d}$ and the ratios
of universal nonperturbative parameters, 
the strange- and nonstrange-quark condensates. 

The content of the paper is as follows: In Sect.~2,
we demonstrate how SU(3)-violation
reveals itself in QCD sum rules. As a study case  we discuss 
the $f_K/f_\pi$ ratio estimated from two-point 
QCD(SVZ) sum rules. In Sect.~3 
we employ light-cone sum rules (LCSR) and
update some previous calculations 
obtaining the differences between the relevant $B\to P$ ($B_{u,d}\to\pi$,
$B_{u,d}\to K$, $B_s\to K$) form factors.
In Sect.~4  we comment on the heavy-mass limit of the SU(3) violation
effects in heavy-to-light form factors.
Sect.~5 contains the analysis of nonfactorizable
corrections in $B\to P_1P_2$ with kaons and pions, employing
LCSR and QCD factorization. 
In Sect.~6, we calculate
the parameter $\delta_{SU(3)}$ in the relation (\ref{rel1})
and analyse two other SU(3)-relations.

\section{The $f_K/f_\pi$ ratio
from SVZ sum rules}

We begin by reminding how the decay constants of pseudoscalar mesons
are calculated  from QCD sum rules \cite{SVZ}.
Comparing the sum rules  for $f_K$ and $f_\pi$
allows to quantify the  SU(3) violation.

In the case of
the pion, the starting point is the correlation function
\begin{eqnarray}
\Pi_{\mu\nu}^{(\pi)}(q)=i
\int d^4xe^{iqx}\langle 0\mid T\{j^{(\pi)}_\mu(x)j^{(\pi)\dagger}_\nu(0) \}\mid
0\rangle
\nonumber
\\
= -\Pi_1^{(\pi)} (q^2)g_{\mu\nu} +\Pi_2^{(\pi)} (q^2)q_\mu q_\nu,
\label{corrpi}
\end{eqnarray}
of two axial-vector quark currents
$j^{(\pi)}_\mu=\bar{u}\gamma_\mu\gamma_5 d$. We use the standard
definition of the pion decay constant,
$\langle 0 | j^{(\pi)}_\mu |\pi(q)\rangle = iq_\mu f_\pi$.

One possible way to obtain $f_\pi$ is to employ the invariant function
\begin{equation}
\Pi^{(\pi)}(q^2)\equiv -\frac{q^\mu q^\nu}{q^2}\Pi_{\mu\nu}^{(\pi)}=
\Pi_1^{(\pi)}(q^2)-q^2\Pi_2^{(\pi)} (q^2)\,,
\label{piampl}
\end{equation}
and write down the  dispersion relation for it:
\begin{equation}
-\Pi^{(\pi)}(q^2)= \frac{f_\pi^2m_\pi^2}{m_\pi^2-q^2}+
\sum\limits_{\pi^{\prime}}\frac{f_{\pi^{\prime}}^2m_{\pi^{\prime}}^2}{m_{\pi^{\prime}}^2-q^2}\,.
\label{piondisp1}
\end{equation}
The r.h.s. contains
the ground-state pion contribution
proportional to $f_\pi^2$; the sum over
$\pi'$  represents, in a simplified form, the
dispersion integral over the excited states with the pion quantum numbers.
Note that the axial meson  $a_1(1260)$ and other hadronic states
with $J^P=1^+$ do not contribute to Eq.~(\ref{piondisp1}).
The amplitude $\Pi^{(\pi)}(q^2)$
is calculated from Eq.~(\ref{corrpi}) using $\partial_\mu j^{(\pi)}_\mu=i(m_u+m_d)\bar{u}\gamma_5d$
and employing the standard tools of current algebra.
At $O(m_{u,d})$ only the contact term proportional
to the quark condensate contributes:
\begin{equation}
\Pi^{(\pi)}(q^2)= -\frac{(m_u+m_d)\langle 0 |\bar{u}u +\bar{d}d|0\rangle}{q^2}
+ O(m_q^2)
\label{contact}
\end{equation}
In order to match Eqs.~(\ref{piondisp1}) and (\ref{contact}),
one has to admit that the decay constants of excited states decouple
in the chiral limit ($f_{\pi'}\sim m_q$). As a result
the well-known Gell-Mann-Oakes-Renner relation \cite{GOR} is reproduced:
\begin{equation}
f_\pi^2 m_\pi^2=-(m_u+m_d)\langle 0 |\bar{u}u +\bar{d}d|0\rangle +O(m_q^2)\,.
\label{GORpi}
\end{equation}
The analogous relation for the kaon is obtained by replacing $d\to s$
everywhere in this derivation:
\begin{equation}
f_K^2 m_K^2=-(m_u+m_s)\langle 0 |\bar{u}u +\bar{s}s|0\rangle
+O(m_s^2)\,.
\label{GORK}
\end{equation}

It is important that the light-quark masses are
independently extracted from various QCD sum rules.
Knowing the value of $m_u+m_d$ one
calculates the nonstrange quark condensate density
from Eq.~(\ref{GORpi}).
We take
\begin{equation}
\langle \bar{q}q\rangle\equiv \langle 0 |\bar{u}u|0\rangle
\simeq \langle 0 |\bar{d}d|0\rangle=-(240 \pm 10 ~\mbox{MeV})^3
\label{cond}
\end{equation}
in the isospin symmetry limit and at the renormalization scale $\mu=1$ GeV.
In what follows we adopt the chiral limit for the $u,d$ quarks
having in mind that $m_{u,d} \ll m_s$.
The interval for the strange quark mass is taken
\begin{equation}
m_s(1 \mbox{GeV})=  130 \pm 20 ~\mbox{MeV}\,,
\label{ms}
\end{equation}
corresponding to $m_s(2~\mbox{GeV})= 100\pm 15 $ MeV,
obtained in the two recent sum rule
analyses \cite{msrecent},
in a good agreement with the lattice QCD results and a recent
determination from $\tau$ decays \cite{Gamiz:2002nu}.
For the strange/nonstrange condensate ratio we adopt
\begin{equation}
\langle \bar{s}s \rangle=(0.8\pm 0.3) \langle \bar{q}q \rangle\,,
\label{ssbar}
\end{equation}
in accordance with the early sum rule analyses for strange baryons
\cite{ssbar}.
This interval also agrees with  more recent estimates \cite{recentssbar}.
We assume that the intervals in Eqs.~(\ref{ms}) and (\ref{ssbar})
are independent from each other
\footnote{
In QCD the ratio of
strange and nonstrange condensates should be correlated with
the mass difference of $s$ and $u,d$ quarks. However,
it is difficult to trace this correlation within the current accuracy of
the sum rules used to estimate these input parameters.}.
It is well known that a numerical comparison of
the two sides in Eq.~(\ref{GORK}) reveals
a rather large $O(m_s^2)$ correction to the r.h.s. (for a recent
analysis, see e.g.\cite{Jamin:2002ev}).
Importantly, the latter correction
can also be estimated  using QCD sum rules for the
correlation function $\Pi^{(K)}$ at the $O(m_s^2)$ level \cite{OvchPivov,Kim}.
The calculated $O(m_s^2)$ terms bring r.h.s. of Eq.~(\ref{GORK}) to an
agreement with the experimental value of its l.h.s..

\begin{figure}[h]
\begin{center}
\includegraphics[height=0.3\textwidth,angle=0]{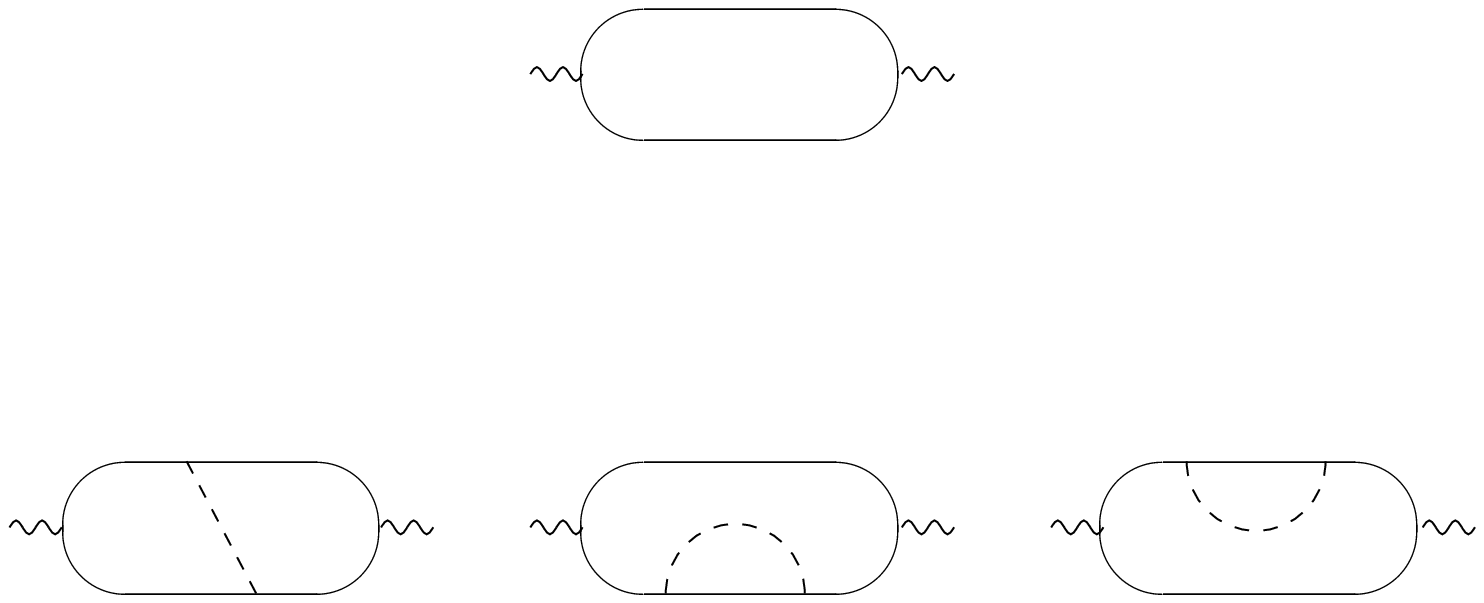}\\
(a)\\
\vspace{1cm}
\includegraphics[height=0.33\textwidth,angle=0]{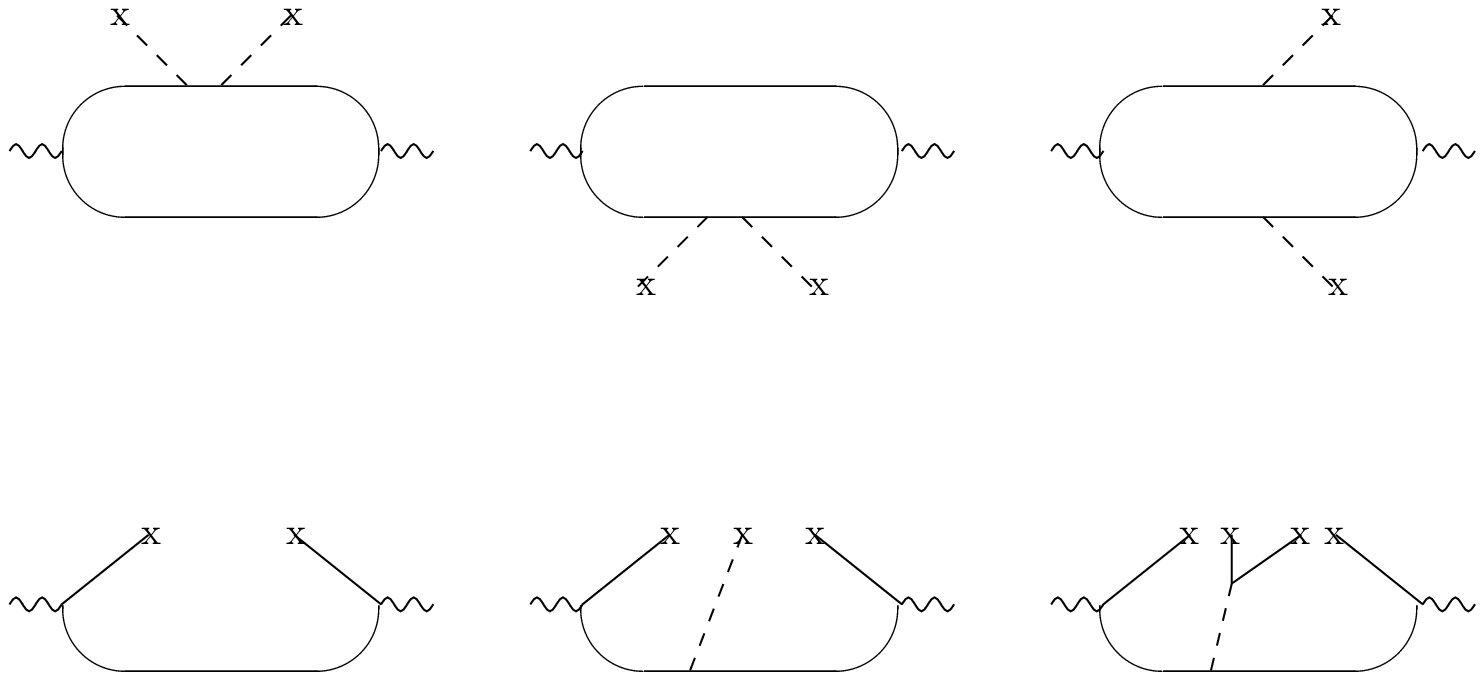}\\
(b)\\
\end{center}
\caption{
{\it  The diagrams corresponding to the OPE for the 
correlation function (\ref{corrpi}): (a) the loop and $O(\alpha_s)$
corrections; (b) the condensate contributions. }}
\end{figure}

In this paper we use an alternative way to calculate
$f_K$ and $f_\pi$, employing    
QCD (SVZ) sum rules \cite{SVZ} derived from the invariant
amplitude $\Pi_2$ in Eq.~(\ref{corrpi}). Taking into account 
the condensates up to dimension 6 and subtracting the sum rule for
$f_\pi^2$ from the one for $f_K^2$ one obtains for the ratio:
\begin{eqnarray}
\frac{f_K^2}{f_\pi^2}= \exp\left({\frac{m_K^2-m_\pi^2}{M^2}}\right)\Bigg\{ 1+ 
\Bigg(\frac{M^2}{4\pi^2f_\pi^2}\Big[
\exp\left(-\frac{s_0^\pi}{M^2}\right)-
\exp\left(-\frac{s_0^K}{M^2}\right)
\Big]\Big(1+\frac{\alpha_s(M)}{\pi}\Big) 
\nonumber
\\
+ \frac{m_s\langle \bar{s}s\rangle - m_d\langle
\bar{q}q\rangle}{f_\pi^2M^2}
+ \frac{16\pi\alpha_s(M)}{81 f_\pi^2 M^4}\left(
9\langle \bar{q}q\rangle\langle \bar{s}s\rangle
+\langle \bar{s}s\rangle^2-10\langle \bar{q}q\rangle^2\right)
\Bigg) \exp\left(\frac{m_\pi^2}{M^2}\right)\Bigg\},
\label{fKfpiratio}
\end{eqnarray}
where  the $O(m_s^2)$  effects are neglected.
In this approximation the gluon-condensate contributions cancel
in the difference of two sum rules, 
and the quark-gluon condensate terms vanish. 
In the above relation the duality threshold parameter 
$s_0^\pi=0.7$ GeV$^2$ and the range of the Borel parameter
$0.5 <M^2 < 1.2 $ GeV$^2$ are fixed from the SVZ sum rule
for the pion decay constant \cite{SVZ}. 
The corresponding parameter for the kaon, $s_0^K$,
is fitted, to achieve the maximal stability of 
the r.h.s. in Eq.~(\ref{fKfpiratio}). 
We obtain $s_0^K=1.05 \mp 0.1~ GeV^2$.
In Fig.~2 the ratio $f_K/f_\pi$   
is plotted, quite stable with respect to $M^2$ and in a good 
agreement with experiment.
\begin{figure}[t]
\begin{center}
\includegraphics[height=0.7\textwidth,angle=0]{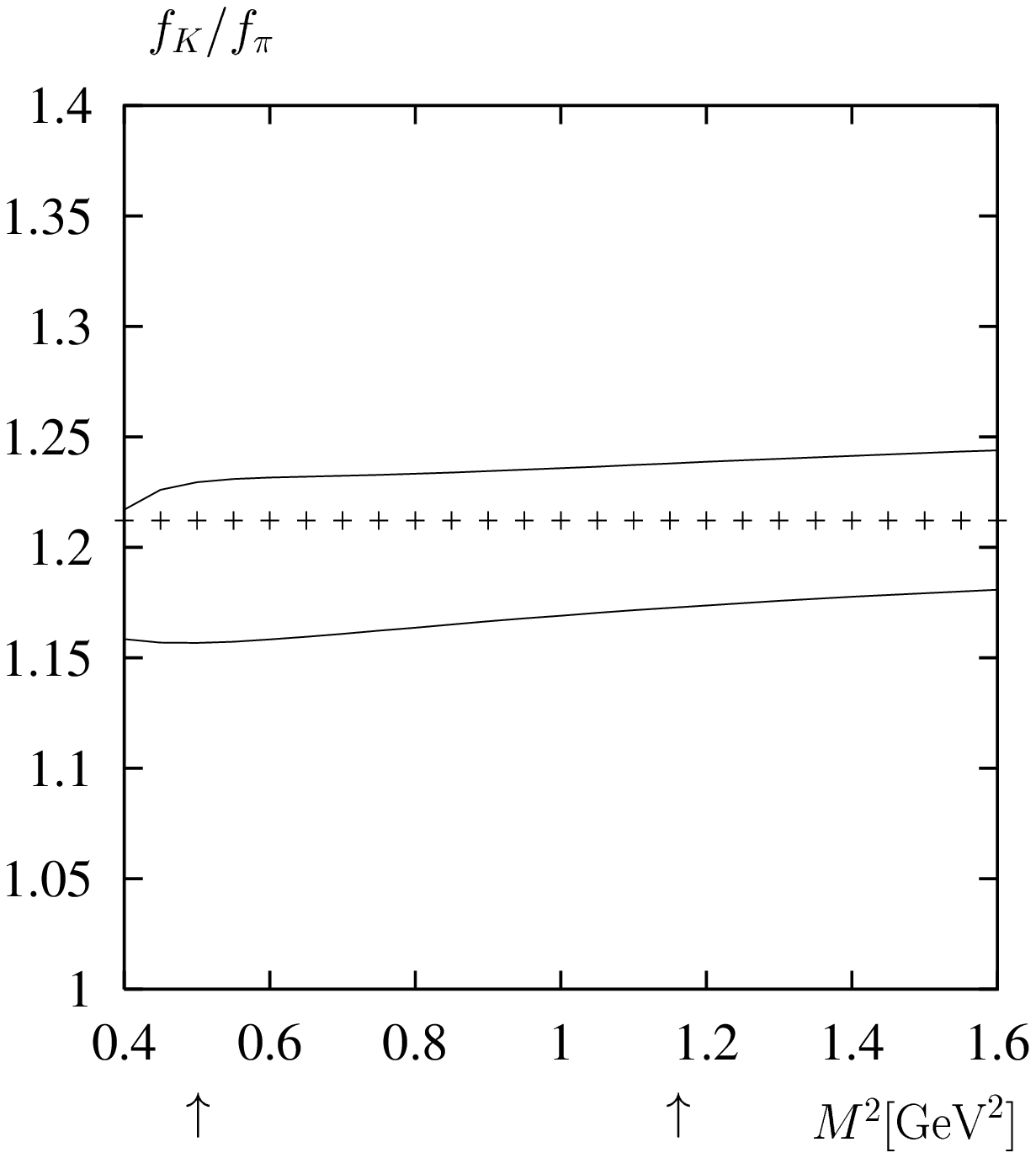}\\
\end{center}
\caption{
{\it  The ratio $f_K/f_\pi$ calculated  
from QCD sum rule (\ref{fKfpiratio}) as a function of 
the Borel parameter,
in comparison with the experimental value
(crosses).  
The upper and lower solid curves 
indicate the interval of theoretical uncertainties.
The arrows indicate the relevant interval of $M^2$. }}
\end{figure}
As expected, the resulting interval $f_K/f_\pi=1.20\pm 0.04$  
is mainly caused by the
uncertainties in $m_s$ and $\langle \bar{s}s \rangle/\langle \bar{q}q
\rangle $. The sum rule relation (\ref{fKfpiratio})
can be further improved by including higher powers
of the $s$ quark mass in the sum rule for $f_K^2$. To give an impression
of their magnitude we write down the complete answer 
for the loop diagram  in this sum rule:
\begin{equation}
[f_K^2]_{loop} = \frac{1}{4\pi^2}
\int_{m_s^2}^{s_0^K}e^{(m_K^2-s)/M^2}
\left(1-\frac{3 m_s^4}{s^2}+\frac{2 m_s^6}{s^3}\right)ds\,.
\label{loop}
\end{equation}
Interestingly, the main contribution to the ratio (\ref{fKfpiratio})
originates from the difference in the threshold parameters for the kaon
and pion channel, whereas the quark-condensate term contributes 
with about 40 \%. The 4-quark condensate contribution (factorized \cite{SVZ} 
into the square of the quark condensates) is small. 
Note that 
parametrically $s_0^K\simeq s_0^\pi+2\sqrt{s_0^\pi}m_s$, 
i.e., the difference between the threshold parameters is of $O(m_s)$.
One can easily expand the ratio (\ref{fKfpiratio}) in  SU(3)-violating
quantities 
$m_s$ and $\langle \bar{s} s \rangle - \langle \bar{q}q \rangle $ obtaining: 
\begin{eqnarray}
f_K/f_\pi \simeq 1+
m_s\Big[\frac{\sqrt{s_0^\pi}}{4\pi^2f_\pi^2}e^{-{s_0^\pi}/M^2}
\left(1+\frac{\alpha_s(M)}{\pi} \right) - \frac{\langle
\bar{q}q\rangle}{2f_\pi^2M^2}\Big]+
\\
\nonumber
+\Big(\langle\bar{q}q\rangle -\langle
\bar{s}s\rangle\Big)
\frac{88\pi\alpha_s\langle\bar{q}q\rangle}{81M^4 f_\pi^2} + O(m_s^2)\,.
\label{msexpand}
\end{eqnarray}

The above analysis clearly demonstrates that QCD sum rules directly relate
the ratio $f_K/f_\pi$ with the differences between strange and nonstrange quark masses and 
condensates. This example justifies the use of sum rules for other 
$SU(3)$-violating ratios considered below.

\section{SU(3) violation in heavy-to-light form factors from LCSR}

To obtain the factorizable part of a given
$B\to P_1P_2$ amplitude ($B=B_{u,d,s}$; $P_{1,2}=\pi,K$)
one needs, in addition to $f_\pi$ and $f_K$,
the $B\to P$ form factors
at the momentum transfer squared $q^2=m_\pi^2\simeq 0$ or $q^2=m_K^2$.
We define these form factors  in a standard way:
\begin{equation}
\langle P(p)|\bar{u} \gamma_\mu b |B(p+q)\rangle=
f^+_{B P}(q^2)\left[(2p +q)_\mu -\frac{m_B^2-m_P^2}{q^2}q_\mu\right]+
f^0_{B P}(q^2)\frac{m_B^2-m_P^2}{q^2}q_\mu\,.
\end{equation}
In the isospin symmetry limit
there are only three flavour combinations:
$B_{u,d}\to \pi$, $B_{u,d}\to K$ and $B_s\to K$. Hereafter
we drop the flavour index $u,d$  at $B_{u,d}$ retaining it only for $B_s$.
The method of QCD light-cone sum rules (LCSR)
\cite{lcsr,BH1,CZB} is used to calculate
the heavy-to-light form factors including SU(3)-violating effects.
Here we will concentrate on the latter aspect of this calculation.
Recent LCSR determinations of $f^+_{B \pi}(q^2)$ can be found in
\cite{KRWWY,BZ}, $f^0_{B\pi}$ was calculated in \cite{KRWinh,BallBK},
 $f^+_{B K}$ in \cite{BKR,BallBK,KRWWY}, and   $f^+_{B_s K}$ in
\cite{LiBsK}.

Let us recall the basic steps of the LCSR derivation.
The correlation function used to calculate the $B\to \pi$ form factors
is
\begin{eqnarray}
F_\mu (p,q)=
i\!\! \int \!d^4xe^{iqx}\langle \pi^+(p)\!\mid T\{\bar{u}\gamma_\mu b(x),
m_b\bar{b}i\gamma_5 d(0)\}\mid\! 0\rangle
\nonumber
\\
= p_\mu F((p+q)^2,q^2)+
q_\mu\widetilde{F}((p+q)^2,q^2)\,.
\label{eq:lcsrcorr}
\end{eqnarray}

At large spacelike $(p+q)^2$ and at $q^2\ll m_b^2$ the operator-product
expansion (OPE) around the
light-cone is used for the product of two currents
in Eq.~(\ref{eq:lcsrcorr}).
The virtual heavy-quark fields are contracted
whereas the light quarks form the light-cone distribution amplitudes
(DA) of the pion, e.g.,  the lowest twist-2 pion  DA
defined in a standard way:
\begin{equation}
\langle \pi^+(p)\mid\bar{u}(x)\gamma_\mu\gamma_5 d(0) \mid 0 \rangle
=-ip_\mu f_\pi \int\limits^1_0 du \,e^{iupx}~\varphi_\pi(u)\,.
\label{pion}
\end{equation} 
The sum rule for $f_{B\pi}^+(q^2)$ is obtained by equating the OPE
result for the invariant amplitude $F$ to the dispersion relation in 
the $B$-meson channel:
\begin{equation}
F((p+q)^2,q^2)=
\frac{2f_B f^+_{B\pi}(q^2)m_B^2}{m_B^2-(p+q)^2}
+\sum\limits_{B_h}\frac{2f_{B_h} f^+_{B_h\pi}(q^2)m_{B_h}^2}{m_{B_h}^2-(p+q)^2}~,
\label{eq:displcsr}
\end{equation}
where the ground-state contribution contains the form factor
multiplied by the $B$-meson decay constant $f_B$. The remaining 
standard steps of the derivation are: the quark-hadron duality
approximation for the sum over
higher states in Eq.~(\ref{eq:displcsr})
and the Borel transformation $(p+q)^2\to M'^2$.
The resulting LCSR reads:
\begin{eqnarray}
f^+_{B\pi}(q^2)=\frac{f_\pi m_b^2}{2m_B^2f_B}
\int\limits_{u_0}^1 \frac{du}u \exp
\left(\frac{m_B^2}{M'^2}-\frac{m_b^2-q^2\bar{u}}{u M'^2}\right)
\Bigg(\varphi_\pi(u,\mu) \nonumber
\\
+ \frac{\mu_\pi}{m_b}
\Bigg[u\varphi_{p}^{(\pi)}(u,\mu) + \frac{\varphi_{\sigma }^{(\pi)}(u,\mu)}{3}
-\frac{u\varphi_\sigma^{(\pi)\prime} (u,\mu)}{6}\Bigg]\Bigg) + ...\,,
\label{eq:fBpisr}
\end{eqnarray}
where $\bar{u}=1-u$,
$\varphi_\sigma^{(\pi)\prime} (u)=d\varphi_\sigma^{(\pi)} (u)/du$,
$u_0=(m_b^2-q^2)/(s_0^B-q^2)$ and
$s_0^B$ is the duality-threshold parameter in the $B$ channel.
The typical values  of the Borel parameter are $M'^2\sim m_B^2-m_b^2$,
the same for the normalization scale $\mu$.
The twist 3 DA's $\varphi_{p,\sigma}$
are normalized with $\mu_\pi=m_\pi^2/(m_u+m_d)$, nonvanishing
in the chiral limit. Additional twist 3 contributions
of quark-antiquark-gluon DA, twist 4 effects \cite{BBKR}
and $O(\alpha_s)$ corrections
\cite{alphas} are not shown in the above expression
but will be taken into account in the numerical
calculation.

For the $B\to K$ form factor, one has to simply adjust the
quark flavours in the correlation function (\ref{eq:lcsrcorr})
replacing $u\to s$
in the vector heavy-light current.
Accordingly, the sum rule for $f^+_{BK}$ is obtained
from Eq.~(\ref{eq:fBpisr}) by replacing
DA's: $\varphi_\pi\to \varphi_K$, $\varphi_{p.\sigma}^{(\pi)}
\to \varphi_{p.\sigma}^{(K)}$, etc.
In addition there are trivial ``kinematical effects'' caused by
the shift of the variable $p^2=m_\pi^2 \simeq 0\to p^2=m_K^2$,
yielding very small $O(m_K^2/m_b^2)$ variations in the exponent and in the
threshold  $u_0$ in Eq.~(\ref{eq:fBpisr}).  
Effects of the same order originate from the variation of the momentum
transfer from $q^2=0$ to $q^2=m_K^2$.

Similarly, the correlation function for the $B_s\to K$ transition 
is obtained by  replacing $d\to s$  
in the pseudoscalar heavy-light current in Eq.~(\ref{eq:lcsrcorr}). 
In this case one also has to replace
$m_B\to m_{B_s}$ and $f_{B}\to f_{B_s}$.
Note that the 2-point sum rule calculation of the $B$ decay  constants
includes  SU(3)-violation, similar to the case
of $f_K/f_\pi$. We will use the most recent estimate \cite{JaminLangefBs}:
\begin{equation}
\frac{f_{B_s}}{f_B}= 1.16\pm 0.05,
\label{eq:ratiofBfs}
\end{equation}
where the uncertainty originates mainly from
$\langle \bar{s}s\rangle/\langle \bar{q}q\rangle$
and $m_s$.  

In the following, we will discuss the SU(3) violation in LCSR 
caused by the differences between the kaon and pion DA's.
It is possible to classify and estimate 
these effects by expanding DA's in the asymptotic and nonasymptotic
parts. One then uses two-point
QCD sum rules to calculate the relevant nonperturbative parameters
entering these expansions. The latter  
include the normalization factors and coefficients of 
the nonasymptotic terms at a low normalization scale.
The twist-2 DA normalization factors are simply $f_\pi$ and $f_K$,
so that one does not need a new calculation. 
The twist-2 pion DA defined in Eq.~(\ref{pion})
is symmetric with respect to $u\to \bar{u}$
transformation (in the isospin limit), and the expansion goes 
over the even Gegenbauer polynomials:\\
\begin{equation}
\varphi_\pi(u,\mu)= 6u(1-u)\Big[1+\sum\limits_{n=2,4,6,...}a_{n}^\pi(\mu)C_{n}^{3/2}(2u-1)\Big]\,,
\label{eq:phipi}
\end{equation}
whereas the kaon DA contains also the odd polynomials
\begin{equation}
\varphi_K(u,\mu)= 6
u(1-u)\Big[1+a_1^K(\mu)C_1^{3/2}(2u-1)+
\sum\limits_{n=2,3,4,..}a_{n}^KC_n^{3/2}(2u-1)\Big]\,. 
\label{eq:phiK}
\end{equation}
In the convention adopted here, $u$ is the longitudinal 
momentum fraction of the strange quark in the kaon.

The coefficient $a^K_{1}$
is related to the difference between the average
momentum fractions
of $s$ and $\bar{d}$($\bar{u}$) quarks in $\bar{K}^{0}$($K^-$):
$a_1^K=5/3\langle x_s-x_{u,d}\rangle=5/3\int _0^1 du (2u-1)\varphi_K(u)$.
The parameter $a_1$ was originally estimated \cite{CZ} using
2-point sum rules for the kaon-interpolating currents. Recently, the
sum rule based on the nondiagonal correlator of pseudoscalar
and axial-vector currents was reanalysed in \cite{BallB} where
a sign error in the previous answer \cite{CZ} for the loop-diagram was found and the
important $O(\alpha_s)$ correction was calculated.
We will use the numerical estimate obtained in \cite{BallB}
\footnote{ We have checked that the signs found in \cite{BallB} are indeed
correct. Note that according to this result the sign of the asymmetry is negative,
opposite to the naive expectation for the heavier $s$-quark to have,
in average, a larger longitudinal momentum fraction.
To finally establish this important parameter of the kaon DA it is desirable to
recalculate it with the same accuracy as in \cite{BallB} also
from the diagonal correlator of the two axial-vector currents, a study
which is beyound the scope of this work. So far,
only the quark-condensate term of the diagonal sum rule is known \cite{CZ}
yielding a positive sign for $a_1$.}:
\begin{equation}
a_1^K(1 \mbox{GeV})= -0.18\pm 0.09\,.
\label{a1sr}
\end{equation}

In our numerical analysis the asymptotic
DA is taken for $\varphi_\pi$.
In order to investigate the uncertainties caused
by possible nonasymptotic effects we allow
for a nonvanishing coefficient $a_2^\pi$ .
With this simple ansatz,
the comparison \cite{BK} of the LCSR for the pion e.m. form factor
with experiment yields the interval
$0 <a_2^\pi(1 GeV)< 0.4$. In order to estimate   the corresponding
$a_2^K$, we use the relation \cite{CZ,BallB} obtained
by subtracting the QCD sum rule for  $a_2^\pi$
from the one for   $a_2^K$ (neglecting the $O(\alpha_s)$ parts) :
\begin{eqnarray}
a_2^K= \frac{e^{m_K^2/M^2}}{f_K^2}
\Bigg[ a_2^\pi f_\pi^2 +
\frac{14}3\Bigg( \frac{m_s\langle \bar{s}s\rangle}{2M^2}-
\frac{5m_s\langle \bar{s}\sigma_{\mu\nu}G^{\mu\nu} s\rangle}{12M^4}
\nonumber
\\
+\frac{8\pi\alpha_s}{27M^4}\left[
3 \langle \bar{q}q\rangle \langle \bar{s}s\rangle
-5\langle \bar{q}q\rangle ^2+2\langle \bar{s}s\rangle^2\right]
\Bigg)\Bigg]\,.
\label{a2K}
\end{eqnarray}
In the above  $G^{\mu\nu} \equiv g_s G^{a\mu\nu}\frac{\lambda^a}{2}$.
The input  is the same as in the sum rule for
$f^2_K/f^2_\pi$ considered in Sect.~2, in addition only the
quark-gluon condensate densities have to be specified. For them we adopt:
\begin{eqnarray}
\langle \bar{q}\sigma_{\mu\nu}G^{\mu\nu} q\rangle =
\left[(0.8 \pm 0.2)\mbox{GeV}^2\right]\langle \bar{q}q\rangle
\,,~~
\frac{\langle \bar{s}\sigma_{\mu\nu}G^{\mu\nu} s\rangle}
{\langle \bar{q}\sigma_{\mu\nu}G^{\mu\nu} q\rangle}=
\frac{\langle \bar{s}s\rangle}{\langle \bar{q}q\rangle}\,.
\end{eqnarray}

The sum rule (\ref{a2K}) yields for the above interval of $a_2^\pi$ :
\begin{equation}
-0.11<a_2^{K}< 0.27\,.
\end{equation}
which includes the interval obtained in \cite{BallB}.
Note that according to the sum rules
the SU(3)-symmetry breaking generates a
nonasymptotic part of the kaon DA (both $a_{1,2}\neq 0$) even
if the pion DA is purely asymptotic.

Concerning higher twist DA's entering LCSR
\footnote{
The complete set of the twist 3,4 DA's
of pseudoscalar mesons worked out in \cite{BraunH,BallDA}
can be found, e.g. in the Appendix B of \cite{BK}.} we
first determine  the normalization factors.
The twist 3 quark-antiquark DA's
$\varphi_{p,\sigma}^{(\pi)}$ and $\varphi_{p,\sigma}^{(K)}$
are normalized by
$\mu_\pi=m_\pi^2/(m_u+m_d)=-2\langle \bar{q}q\rangle/f_\pi^2 $
and $\mu_K=m_K^2/(m_u+m_s)$ , fixed by our choice of the quark condensate
density and $m_s$, respectively. The remaining
input parameters are the normalization factors $f_{3\pi,3K}$ and
$\delta_{\pi,K}^2$ of the
twist-3 quark-antiquark-gluon and twist-4 DA's, respectively,
as defined in \cite{BraunH,BallDA}.
We use $f_{3\pi}=0.0035 $ GeV$^2$ and $\delta_\pi^2=0.17 \pm 0.05$
GeV$^2$ determined from the two-point
QCD sum rules \cite{NSVVZ,CZwf,CZ}.
To assess the level of SU(3)-violation in these parameters
we present in the Appendix a new sum rule
calculation of $\delta_K^2$, yielding
\begin{equation}
\frac{\delta_K^2 f_K}{\delta_\pi^2 f_\pi}=1.07^{+0.14}_{-0.13}\,.
\end{equation}
For $f_{3K}$, the
sum rule calculation is more complicated and we postpone
it to the future. Having in mind the result above, we assume:
\begin{equation}
\frac{f_{3K}}{f_{3\pi}}= 1.0 \pm 0.2\,.
\end{equation}
We also adopt purely asymptotic
higher twist DA's, in particular we neglect possible
asymmetries in the kaon twist 3,4  DA's analogous to $a_1^K\neq 0$.
At the same time,  we take into account
the mass corrections to the twist 3,4 kaon DA's
\cite{BallDA}, due to the mixing of various twists
at the $O(m_K^2)$ level.

Having specified the DA parameters
we are able to calculate the form factors numerically,
using LCSR (\ref{eq:fBpisr}) and the analogous sum rules
for $B\to K$ and $B_s\to K$ form factors.
The remaining input parameters are the same as
in \cite{KRWWY}: $m_b=4.7\pm 0.1$ GeV (the one-loop $b$-quark pole
mass), $s_0^B=35  \mp 2$ GeV$^2$ and $M'^2= 8\mbox{-} 12$ GeV$^2$.
The normalization scale is $\mu_b=m_B^2-m_b^2$.
With the above input we predict $f^+_{B\pi}(0)=0.25^{+0.05}_{-0.02}$,
an interval close to the ones obtained in \cite{KRWWY,BZ}.
Simultaneously, the following ratio of the
$B\to K$ and $B\to \pi$ form factors is obtained:
\begin{equation}
f^+_{BK}(0)/f^+_{B\pi}(0)=1.08^{+0.19}_{- 0.17}\,,
\label{Bkbpi}
\end{equation}
where the separate uncertainties due to the spread of
the independent input parameters are added in quadrature.
Note that the $s$-quark mass and the condensate-ratio
dependence of all input parameters in LCSR is taken into account
in a correlated way.
Numerically, the SU(3) violation effect originates mainly from
the ratio of the twist 2 normalization factors $f_K/f_\pi$ and from the asymmetry
$a_1^{K}\neq 0$. Both quantities are
calculable from 2-point sum rules, as we have seen above.
We can thus trace the origin of the ratio (\ref{Bkbpi})
to $m_s$ and the ratio of strange and nonstrange condensates.
Moreover, the uncertainty of our predictions is to a large extent
due to the variation of $m_s$ and of the condensate ratio.
The remaining uncertainties
in both sum rules, such as the ones caused by the intervals of $m_b$
and $M'^2$ largely cancel in the ratio.

Turning to the $B_s\to K$ -transition, we note that
here the strange-nonstrange asymmetry in the kaon DA
has effectively an opposite sign to the $B\to K$ case,
because the $s$ quark is now a ``spectator''.
In other words, we can use in LCSR the same DA $\varphi_K(u)$ but
with $a_1^K$ having an opposite sign. We obtain
\begin{equation}
f^+_{B_s K}(0)/f^+_{B\pi}(0)=1.40^{+0.12}_{- 0.13}\,,
\label{fBsK}
\end{equation}
quite a substantial effect.
Our numerical results (\ref{Bkbpi}) and (\ref{fBsK}) are different from the ones
obtained earlier in \cite{BKR, BallBK,KRWWY,LiBsK} because of the sign change
of the parameter $a_1$. Note that LCSR predict
substantial magnitudes of SU(3) violation also for
the ratios of the $B\to \rho,K^*,\phi$  form factors \cite{BallBraun}.

In addition, we have checked numerically that the change of the
kinematical variable $p^2$ from zero to $m_K^2$
in the correlation function as well as the switch to the momentum
transfer $q^2=m_K^2$, being both $O(m_K^2/m_b^2)$ are $\leq 1 \%$.
Having in mind uncertainties of our calculation we
neglect the latter small changes  and use in all amplitude relations
$f^0_{BP}(m_K^2)\simeq f^0_{BP}(m_\pi^2)=f^0_{BP}(0)=f^+_{BP}(0)$,
so that the factorizable amplitudes defined in Eq.~(\ref{Afact})
are:
\begin{equation}
A_{fact}(B\to P_1P_2)\simeq
i\frac{G_F}{\sqrt{2}}m_B^2f_{P_2}f^+_{BP_1}(0)\,.
\label{Afact1}
\end{equation}

Finally, using Eqs.~(\ref{Bkbpi}) and (\ref{fBsK}) we predict  SU(3) violation
in the factorizable $B\to P_1 P_2$ amplitudes,
for all possible flavour configurations (in the isospin limit):
\begin{equation}
\left.
\begin{array}{r@{\quad}lc}
A_{fact}(B\to \pi K)&=&\frac{f_K}{f_\pi}A_{fact}(B\to\pi\pi)=1.22\\
&&\\
A_{fact}(B\to K \pi)&=&1.08^{+0.19}_{- 0.17}\\
&&\\
A_{fact}(B\to K \bar{K})&=&1.31^{+0.24}_{- 0.21}\\
&&\\
A_{fact}(B_s\to K\bar{K})&=&1.76^{+0.15}_{- 0.17}\\
&&\\
A_{fact}(B_s\to K \pi)&=&1.45 ^{+0.13}_{- 0.14}\\
\end{array} \right\}\times A_{fact}(B\to\pi\pi)\,.
\end{equation}

We conclude that in certain cases flavour SU(3)
is not a reliable symmetry for charmless $B$ decays.
Instead of using SU(3) relations one should better rely on the
QCD calculation of separate decay amplitudes.

\section{ Heavy quark limit of SU(3) violation}

With the help of LCSR it is possible to study
the $m_b\to \infty$ behaviour of the $B\to P$ form
factors. Making the standard substitutions:
$m_B^2=m_b^2+2m_b\Lambda$, $s_0^B=m_b^2+2\omega_0m_b$,
so that $u_0^B\simeq 1-\omega_0/m_b$,
$M'^2 =2m_b\tau$ and $f_B=m_b^{-1/2}\hat{f}_B$
one extracts the heavy mass scale in all $m_b$-dependent
parameters in the sum rule (\ref{eq:fBpisr}), obtaining
\begin{eqnarray}
\lim_{m_b\to\infty}f^+_{B\pi}(0)=m_b^{-3/2}\Bigg\{\frac{f_\pi}{2\hat{f}_B}exp\left(\frac{\Lambda}{\tau}\right)\int_0^{2\omega_0}
d\rho~ exp\left( -\frac{\rho}{\tau}\right)\Bigg\}
\Big[ -\rho \varphi_\pi^\prime(1)
\nonumber
\\
+\mu_\pi\left(\varphi_p^{(\pi)}(1)-
\frac{\varphi_\sigma^{(\pi)\prime}(1)}6\right)\Big]
+O(m_b^{-5/2})\,.
\end{eqnarray}

Replacing $\pi \to K$  with our choice of twist 2 DA
we get
$
\varphi_\pi\to \varphi_K(u)=6u(1-u)(1+3a_1(2u-1))
$,
with
$
a_1\sim O\left(m_s/M\right)\,
$
and the scale $M\sim 1$ GeV.
We immediately notice that certain SU(3) violating effects
survive  in the ratio $f_{BK}/f_{B\pi}$
at $m_b\to \infty$. The fact that the flavour SU(3)-symmetry
remains broken in the heavy-quark limit
seems quite natural. Even if the
light quarks in the $B\to P$ transition
originate from the decay of a very heavy $b$ quark,
there is always a long-distance part of SU(3)-violation
manifesting itself in the ratios of normalization constants
$f_K/f_\pi$, $\mu_K/\mu_\pi$ and in the asymmetry in the kaon twist-2 DA.

\section{SU(3) violation in nonfactorizable amplitudes}

After having calculated the magnitude of SU(3) violation
in the factorizable $B\to P_1P_2$ amplitudes,
the remaining task is to investigate the SU(3) effects in
the process-dependent nonfactorizable contributions.
We will mainly concentrate on the charmless decay amplitudes entering
the relation (\ref{rel1}).
The effective weak Hamiltonian is given by
\begin{eqnarray}
H_W=\frac{G_F}{\sqrt{2}}\sum\limits_i\lambda_ic_i O_i\,,
\end{eqnarray}
where $\lambda_i$ , $c_i$ and $O_i$ are the CKM factors, Wilson coefficients
and effective operators, respectively.
Each  decay amplitude can be represented as a decomposition
in the hadronic matrix elements of $O_i$ with different
contractions of quark lines (topologies) \cite{BS}:
\begin{eqnarray}
&&A(B\to P_1P_2)\equiv \langle P_1P_2 |H_W| B\rangle \nonumber \\
&&=\sum\limits_{T=E,P,A,..}A_T(B\to P_1P_2)=
\frac{G_F}{\sqrt{2}}\sum\limits_{T=E,P,A,..}\sum_i\lambda_ic_i \langle
P_1P_2| O_i|B\rangle_T\,.
\end{eqnarray}
For  the decay channels involved in the relation (\ref{rel1})
it is sufficient to
consider the hadronic matrix elements of the current-current
operators $O_{1,2}$ in the emission topology. These matrix elements
are the only ones which enter $A(B^-\to \pi^-\pi^0)$.
The additional  annihilation and penguin contributions
to $A(B^-\to \pi^0 K^-)$ cancel  (in the isospin symmetry limit)
with the amplitude $A(B^-\to \pi^-\bar{K}^0)$ which contains
only annihilation  and penguin  terms
(remember that we neglect electroweak penguins), so that
\begin{eqnarray}
\sqrt{2}A(B^-\to \pi^0 K^-)+A(B^-\to \pi^-\bar{K}^0)
=\sqrt{2}A_E(B^-\to \pi^0 K^-)\,.
\end{eqnarray}

The two relevant amplitudes are  given by the following combinations of
hadronic matrix elements:
\begin{eqnarray}
A_E(B^-\to \pi^0 K^-)=\frac{G_F}{\sqrt{2}}
V_{us}V_{ub}^*\Bigg\{
\left(c_1+\frac{c_2}{3}\right)\langle \pi^0 K^-|O_1^{(s)}|B^-\rangle_E
+2c_2\langle \pi^0 K^-|\tilde{O}_1^{(s)}|B^-\rangle_E
\nonumber
\\
+\left(c_2+\frac{c_1}{3}\right)\langle K^- \pi^0|O_2^{(s)}|B^-\rangle_E
+2c_1\langle K^- \pi^0|\tilde{O}_2^{(s)}|B^-\rangle_E\Bigg\}
\nonumber
\\
=\frac{V_{us}V_{ub}^*}{\sqrt{2}}\Bigg[A_{fact}(B\to \pi K)\left(c_1+\frac{c_2}{3}+2c_2r^{B\pi
K}_E\right)+
A_{fact}(B \to K \pi)\left(c_2+\frac{c_1}{3}+2c_1r^{BK\pi}_E\right)\Bigg],
\label{decomp1}
\end{eqnarray}

\begin{eqnarray}
A(B^-\to \pi^-\pi^0)=\frac{G_F}{\sqrt{2}}
V_{ud}V_{ub}^*\Bigg\{
\left(c_1+\frac{c_2}{3}\right)\langle \pi^0 \pi^-|O_1^{(d)}|B^-\rangle_E
+2c_2\langle \pi^0 \pi^-|\tilde{O}_1^{(d)}|B^-\rangle_E
\nonumber
\\
+\left(c_2+\frac{c_1}{3}\right)\langle \pi^- \pi^0|O_2^{(d)}|
B^-\rangle_E
+2c_1\langle \pi^- \pi^0|\tilde{O}_2^{(d)}|B^-\rangle_E
\Bigg\}
\nonumber
\\
=\frac{V_{ud}V_{ub}^*}{\sqrt{2}}
A_{fact}(B\to \pi \pi)\Bigg[\frac{4}3(c_1+c_2)+2(c_1+c_2)r^{B\pi\pi}_E\Bigg]\,,
\label{decomp2}
\end{eqnarray}
where the current-current operators are
$O_1^{(n)}= (\bar{n}\Gamma_\mu u)(\bar{u}\Gamma^\mu b)$, and
$O_2^{(n)}= (\bar{u}\Gamma_\mu u)(\bar{n}\Gamma^\mu b)$, 
($n=s,d$; $\Gamma_\mu=\gamma_\mu(1-\gamma_5$)) and 
we used Fierz transformations:
$O_{1,2}^{(n)}=\frac{1}{3}O_{2,1}^{(n)}+2\tilde{O}_{2,1}^{(n)}$, so
that $\tilde{O}_1^{(n)}= 
(\bar{n}\Gamma_\mu \frac{\lambda^a}{2} u)(\bar{u}\Gamma^\mu
\frac{\lambda^a}{2}b)$ and 
$\tilde{O}_2^{(n)}= (\bar{u}\Gamma_\mu \frac{\lambda^a}{2} u)(\bar{n}\Gamma^\mu
\frac{\lambda^a}{2}b)$ . In the
relations (\ref{decomp1}) and (\ref{decomp2})
we introduced the ratios of 
matrix elements in the emission topology: 
\begin{equation}
r^{(BP_1P_2)}_E =\frac{\langle P_1P_2|\tilde{O}_{i}^{(n)}|B\rangle_E}{\langle P_1P_2|O_{i}^{(n)}|B\rangle_E}\,,
\end{equation}
where $i=1$ or 2 and $P_2$ is the emitted meson. 
In the third lines in Eqs.~(\ref{decomp1}),(\ref{decomp2})
we take into account that, 
in first approximation, 
the matrix elements of $O_{1,2}$ 
coincide with the corresponding factorizable amplitudes.
The matrix elements of $\tilde{O}_{1,2}$ 
accumulate nonfactorizable effects  
originating from  the hard- and 
soft-gluon exchanges.
We will take them into account in
$O(\alpha_s)$ and $O(1/m_b)$, respectively. 
Using the notation introduced in Eq.~(\ref{schem}), we
separate these two effects:
\begin{equation}
r^{(BP_1P_2)}_E=
\frac{\alpha_s C_F}{\pi}\delta^{(BP_1P_2)}_E
+\frac{\lambda_E^{(B P_1P_2)}}{m_B}\,.
\end{equation}
The hadronic matrix elements of $\tilde{O}_{1,2}$ and correspondingly 
the ratios  $r^{BP_1P_2}_E$ are calculable from LCSR 
using the method suggested in \cite{AKBpipi}. 

To exemplify  the LCSR calculation we consider the matrix element
$\langle \pi^+ K^-|\tilde{O}_1^{(s)}|\bar{B}^0\rangle_E=r^{(B\pi K)}A_{fact}(B\to \pi K)$.
The starting point is the correlation function
\begin{eqnarray}
 F^{(B\pi K)}_\alpha(p,q,k)
  =-\int d^4x\ e^{-i(p-q)x} \int d^4y\ e^{i(p-k)y}
\langle 0 |T\left\{j^{(K)}_\alpha(y)\tilde{O}_1^{(s)}(0)j^{(B)}_5(x)
\right\} |\pi^-(q)\rangle
\nonumber
\\
=(p-k)_\alpha F^{(B\pi K)}+ ...
\label{eq:BKpicorr}
\end{eqnarray}
where $j^{(K)}_\alpha =\bar{u}\gamma_\alpha\gamma_5 s$ and
$j_5^{(B)}=i m_b \bar{b}\gamma_5 d$ are the quark currents
interpolating kaon and $B$ meson, respectively.
We only need the invariant amplitude $F^{(B\pi K)}$ which
depends on the kinematical invariants $(p-q)^2$, $(p-k)^2$ and
$P^2\equiv(p-q-k)^2$, the other amplitudes in Eq.~(\ref{eq:BKpicorr})
are denoted by ellipses. Following the derivation in
\cite{AKBpipi}, one uses dispersion relations, quark-hadron duality
and Borel transformation in both kaon and $B$ meson channels
characterized by the variables $(p-k)^2$ and $(p-q)^2$, respectively.
The variable $P^2$ is analytically continued to the
physical point $m_B^2$, so that the artificial momentum $k$
vanishes in the resulting LCSR for the hadronic matrix element:
\begin{eqnarray}
&&\langle K^-(p)\pi^+(-q)|\tilde{O}_1|\bar{B}^0(p-q)\rangle
= \frac{-i}{\pi^2 f_B f_K m_B^2}
\int_{m_b^2}^{s_0^B}ds_2~e^{(m_B^2-s_2)/M'^2}
\nonumber
\\
&&\times \int_{m_s^2}^{s_0^K}ds_1 ~e^{(m_K^2-s_1)/M^2}
\mbox{Im}_{s_2}~\mbox{Im}_{s_1}~F^{(B\pi K)}(s_1,s_2,m_B^2)\,.
\label{eq:generalsr}
\end{eqnarray}
The amplitude $F^{(B\pi K)}$  and its imaginary part are calculated
using light-cone OPE in the domain
$(p-k)^2,(p-q)^2,P^2<0,|(p-q)^2|,|(p-q)^2|,|P^2|\gg\Lambda_{QCD}$.
It is important for the consistency of the method that
the factorizable amplitude containing the product of $f_K$
and the LCSR for $B\to \pi$ form factor can be restored \cite{AKBpipi}
from the correlation function similar to Eq.~(\ref{eq:BKpicorr})
but with the operator $O_1^{(s)}$.
The corresponding tree-level diagram is shown in Fig.~3.

\begin{figure}
  \centering
  \includegraphics[trim=0 20 0 0, width=0.4\textwidth]{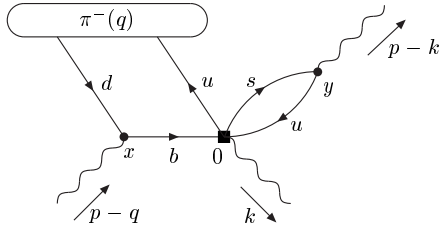}
  \caption{Tree-level diagram corresponding to the
correlation function similar to (\ref{eq:BKpicorr}), with operator $O_1^{(s)}$.}
\end{figure}

The QCD result for  $F^{(B\pi K)}$ is determined by the diagrams
shown in  Fig.~4,5 which contain an additional gluon
exchange that violates factorization.
These diagrams represent convolutions of hard-scattering
amplitudes formed by virtual quarks and gluons at light-cone
separations, with the pion light-cone DA's of growing twist
accumulating the long-distance dynamics.
\begin{figure}
\centering
\subfigure[]{\includegraphics[trim=0 20 0 0, width=0.4\textwidth]{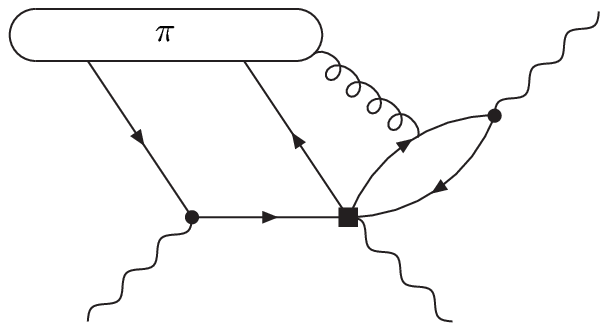}}%
\subfigure[]{\includegraphics[trim=0 20 0 0, width=0.4\textwidth]{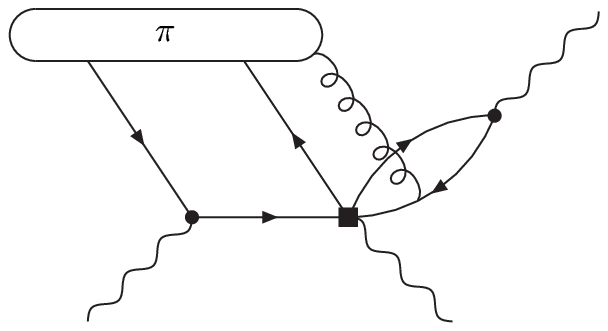}}%
\\
\subfigure[]{\includegraphics[trim=0 20 0 0, width=0.4\textwidth]{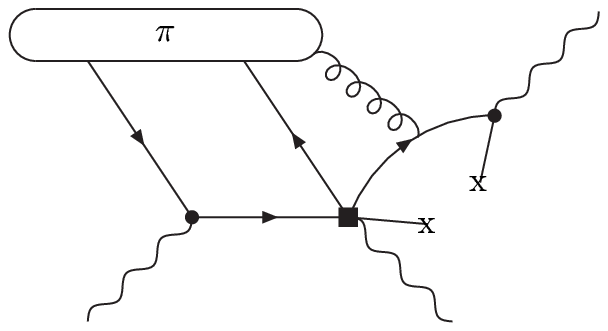}}%
\subfigure[]{\includegraphics[trim=0 20 0 0, width=0.4\textwidth]{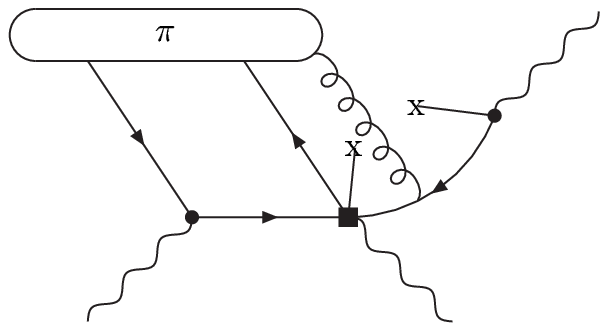}}%
  \caption{Diagrams corresponding to the soft-gluon
contributions in the correlation function (\ref{eq:BKpicorr}).}
\end{figure}

\begin{figure}
\centering
\subfigure[]{\includegraphics[trim=0 20 0 0, width=0.4\textwidth]{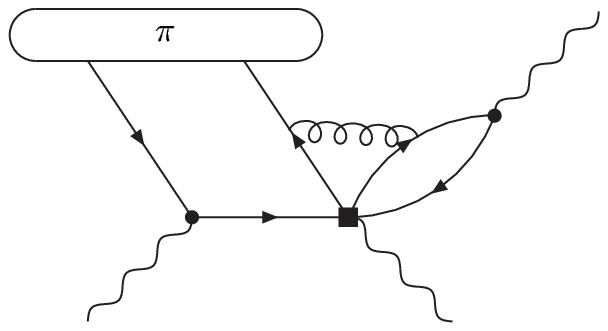}}%
\subfigure[]{\includegraphics[trim=0 20 0 0, width=0.4\textwidth]{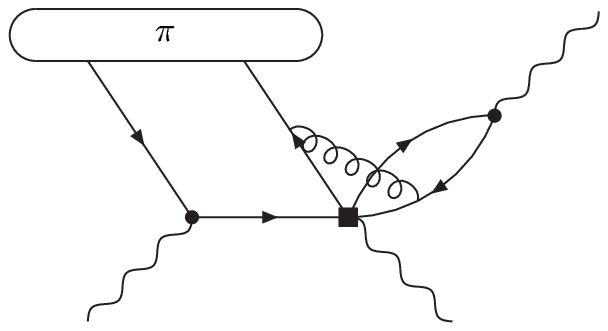}}%
\\
\subfigure[]{\includegraphics[trim=0 20 0 0, width=0.4\textwidth]{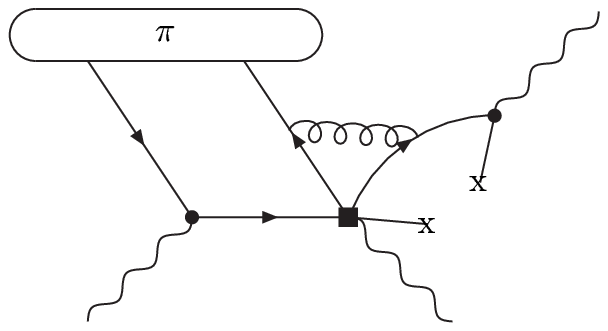}}%
\subfigure[]{\includegraphics[trim=0 20 0 0, width=0.4\textwidth]{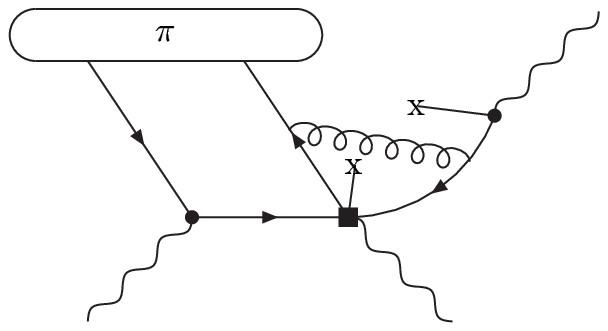}}%
\caption{Some of the diagrams corresponding to the $O(\alpha_s)$
contributions in the correlation function (\ref{eq:BKpicorr}):(a,b) hard-gluon
(c,d) hard-gluon and quark condensate. The similar diagrams where the gluon is attached to
the $b$- and $d$-quark lines are not shown.}
\end{figure}

So far, only the soft-gluon part of the sum rule
for $B\to \pi\pi$
was obtained  \cite{AKBpipi}  resulting in the estimate
for $\lambda_E^{(B \pi\pi)}$.
Here we will extend this calculation
to the channels with kaons in order to obtain
$\lambda_E^{(B\pi K)}$ and $\lambda_E^{(B K\pi)}$.
The soft-gluon
contribution to $B  \to \pi K$ originates from the
diagram in Fig.~4a,b which are similar to the diagrams
determining
the LCSR relation for $\lambda_E^{B\pi \pi}$
obtained in \cite{AKBpipi}.
In addition, for the correlation function (\ref{eq:BKpicorr})
there are new diagrams
shown in Fig.~4b,c   which are absent in the case of $B\to \pi\pi$
(in the chiral limit). These
diagrams correspond to the
four-quark-gluon contributions to the pion DA and are factorized
in terms of the quark condensate and quark-antiquark-gluon DA.
Similar condensate contributions have been taken into account
in LCSR for the penguin matrix elements in $B\to \pi\pi$
\cite{ourpenguins}
where one can find a more detailed discussion. 
The sum rule relation obtained from Eq.~(\ref{eq:generalsr})
reads:
\begin{eqnarray}
&&\lambda^{(B\pi K)}_E= \frac{m_B}{f^{+}_{B\pi}(0)}
\Bigg(\frac{1}{4\pi^2f_K^2}\int\limits _0^{s_0^K}ds~
e^{-s/M^2}\Bigg)
\Bigg(\frac{m_b^2}{2f_Bm_B^4}\int\limits_{u_0^B}^1 
\frac{du}{u^2}e^{\,m_B^2/M'^2-m_b^2/uM'^2}
\nonumber
\\
&&\times
\Bigg[m_bf_{3\pi}\Bigg( 1+\frac{4\pi^2m_s\langle \bar{q}q \rangle}{3M^4}
-\frac{4m_s^2}{M^2}
\Bigg)
\int\limits_0^u \frac{dv}{v} 
\varphi_{3\pi}(1-u,u-v,v)
\nonumber
\\
&&+f_\pi\delta_\pi^2 \Big(1+O(m_s\langle \bar{q}q \rangle)\Big)\widetilde{\varphi}^{tw4}_\pi(u)
\Bigg]\Bigg)\,,
\label{soft}
\end{eqnarray}
where $\varphi_{3\pi}(\alpha_i)=360 \alpha_1\alpha_2\alpha_3^2$ is 
the twist-3 quark-antiquark-gluon DA taken in the asymptotic
form, $f_{3\pi}$ is the corresponding normalization constant.
We have  also taken into account the $O(m_s^2)$ correction to the perturbative
loop and the twist-3 part of the quark-condensate
term.
Since this term turned out to be numerically extremely  small 
we have neglected the corresponding $O(m_s\langle \bar{q}q \rangle)$ 
twist-4 contribution  indicated in Eq.~(\ref{soft}). 
The same argument holds for the corrections of order $m_s^2/M^2$ to the
twist-4 part which we calculated but found to be negligible.
Consequently, in Eq.~(\ref{soft}), $\widetilde{\varphi}^{tw4}(u)$ 
denotes the same combination of twist 4 
quark-antiquark-gluon DA's which enters 
LCSR  for $B\to \pi\pi$, and can be easily read off from Eq.(30) 
in \cite{AKBpipi}. Finally, for  $f^{+}_{B\pi}(0)$ we use LCSR
(\ref{eq:fBpisr}).
Comparing the sum rule for $\lambda^{(B\pi K)}$ with 
the one for $\lambda^{(B\pi\pi)}$
one immediately recognizes that SU(3) violation
originates from the differences in the emitted meson
channels: $f_K$ vs $f_\pi$ , $s_0^K$ vs $s_0^\pi$ and the absence of the
quark-condensate and $O(m_q^2)$ terms in $\lambda^{(B\pi\pi)}$.
In the $B\to K \pi$
channel (with the emitted pion) the SU(3) violation
with respect to $B\to \pi \pi$  has another origin
and is due to  the  differences between the kaon and pion
DA's which were already discussed in the previous section.
Thus, in order to obtain the sum rule for  $\lambda^{(BK\pi)}$
one has to replace in Eq.~(\ref{soft}) $f^+_{B\pi}(0)\to f^+_{BK}(0)$,
$f_{K(\pi)}\to f_{\pi(K)}$,
$s_0^K\to s_0^\pi$ , $m_s\to 0$ (quark condensate terms vanish),
$f_{3\pi}\to f_{3K}$, $\varphi_{3\pi} \to \varphi_{3K}$,
$\delta^2_\pi\to \delta^2_K$, $\tilde{\varphi}_\pi^{tw4}\to \tilde{\varphi}_K^{tw4}$.
Numerically, we obtain:
\begin{equation}
\lambda_E^{(B\pi \pi)}= 110 \pm 40~\mbox{MeV}\,,~~
\lambda_E^{(B\pi K)}= 120^{+34}_{-43}~ \mbox{MeV}\,,~~
\lambda_E^{(B K\pi)} = 109^{+39}_{-45}~ \mbox{MeV}\,,~~
\label{lambda}
\end{equation}
where the uncertainties are correlated. We find that the magnitude of SU(3)-breaking
in $\lambda_E^{(B P_1 P_2)}$ is
generally smaller than in the form factors
revealing that the effects
due to $m_s$ and $\langle \bar{s}s \rangle/ \langle \bar{q}q \rangle $
largely cancel in the ratios of nonfactorizable and factorizable amplitudes.

The two-loop diagrams in Fig. 5 have not been calculated yet,
nevertheless in order to clarify the origin of SU(3) effects
it is sufficient to write down  the answer for these diagrams in
a generic form:
\begin{eqnarray}
\label{hard}
\mbox{Im}_{s_2}~\mbox{Im}_{s_1}~F^{(B\pi K)}(s_1,s_2,m_B^2)^{(Fig.5)}=
\frac{\alpha_sC_F}{\pi}\Bigg[
T_{5a,b}(s_1,s_2,m_b,m_s^2)\\
\nonumber +
m_s\langle\bar{q}q \rangle T_{5c,d}(s_1,s_2,m_b)
\Bigg]\varphi_\pi(s_2/m_b^2)\,,
\end{eqnarray}
where, for simplicity only the leading twist-2 part is shown.
In the above, the indices at the hard amplitudes $T$ denote
the corresponding diagrams. Substituting Eq.~(\ref{hard})
in Eq.~(\ref{eq:generalsr}) we observe that SU(3)-violation
with respect to $B\to \pi\pi$
is again due to the differences in the channel of the emitted meson:
1) $f_K \neq f_\pi$ , $s_0^K \neq s_0^\pi$ ;
2) quark condensate $O(m_s)$ contributions; 3) $O(m_s^2)$ effects.
The analogous expression for $B  \to K \pi$ is obtained
by the following replacements in Eqs.~(\ref{hard}),(\ref{eq:generalsr}):
$\varphi_\pi\to \varphi_K$, $s_0^K\to s_0^\pi$ , $f_K\to f_\pi$, $m_s\to 0$.
As in the case of the soft contribution, now the differences between DA's
of kaon and pion determine the SU(3)-violation.

After this qualitative discussion we still need
to estimate the hard-gluon
contribution numerically. For that we employ
QCD factorization.
The expressions for the matrix elements can be found
in \cite{BBNS} and we will not repeat them here.
As an input in this calculation we use the LCSR form factors,
and adopt the normalization scale $\mu_b$. In addition
we take from \cite{BBNS}
the inverse moments of the $B$ meson DA and of the
pion twist 3 DA: $\lambda_B= 0.35\pm 0.15$ GeV and $X_\pi^H=2.4\pm 2.4$
GeV, respectively.
The numerical result is:
\begin{eqnarray}
\frac{\alpha_s C_F}{\pi}\delta_E^{(B\pi\pi)}
=(-0.025)\mbox{-}(+ 0.044) -0.045i\,,
\nonumber
\\
\frac{\alpha_s C_F}{\pi}\delta_E^{(B\pi K)}
= (-0.035) \mbox{-} (+ 0.032) -(0.040\pm 0.002)i\,,
\nonumber
\\
\frac{\alpha_s C_F}{\pi}\delta_E^{(B K \pi)}
= (-0.029)  \mbox{-} (+0.055) -0.045i\,.
\label{delta}
\end{eqnarray}
The uncertainties in the real parts are due to
the spread in $\lambda_B$ and $a_2^{\pi,K}$, and the small
uncertainty in the imaginary part of $\delta_E^{(B\pi K)}$ is due to
$a_1^K$. Altogether the
uncertainties in the real parts overshoot the ones related to
the SU(3) breaking.
Combining Eqs.~(\ref{lambda}) and (\ref{delta})
we obtain the parameters $r^{(B P_1P_2)}$
that are needed to complete the calculation of the matrix elements
(\ref{decomp1}) and (\ref{decomp2}).

Before closing this section, let us mention that
the LCSR analysis of nonfactorizable contributions can easily be extended
to the matrix elements of the quark-penguin operators $O_{3-6}$ as far as
the emission topology is concerned. Because of the $(V+A)$ structure of the operator
$O_5$ (which becomes $(S+P)$ after Fierz transformation), we expect
the result to change qualitatively:
First, in the chiral limit $m_q\to 0$, the Fig.~4a,b diagrams  have to
vanish due to chiral symmetry. Consequently, the loop diagram
is proportional to $m_s$ if
the emitted particle is a kaon, and vanishes if it is a pion.
Second, due to the changed Dirac structure of the correlator, the leading twist is
4. This implies that the soft-gluon nonfactorizable
correction is suppressed by \(1/m_b^2\).
In total, we get a result of the form
\begin{equation}
\frac{\lambda^{B P_1 K}_{V+A}}{m_B}\sim O\left
(\frac{\Lambda_{QCD}}{m_b^2}\right)\left(O(m_s)
+O\left(\frac{\langle\bar{s}s\rangle-\langle\bar{u}u\rangle}{M^2}\right)\right)\,.
\end{equation}
It is interesting to note that also the quark condensate term vanishes if
the emitted particle is a pion, as long as we rely on isospin symmetry.
We postpone a more detailed study of these
contributions, as well as the analysis of the SU(3)-violation in the
penguin-topology contributions (generated by current-current
and penguin operators) to the future. In fact, some results can already be
read off from the LCSR estimates for
gluonic penguins and charming penguins \cite{ourpenguins} replacing
pions by kaons. However,
in  most of $B\to PP$ decay amplitudes,  the penguin effects
are accompanied  by annihilation contributions. The latter
have not yet been analysed within the LCSR approach. The
annihilation amplitudes with hard-gluon exchanges are
also problematic for the QCD factorization approach. Therefore
the uncertainties caused by annihilation effects are at the moment
certainly larger than any SU(3)-breaking in the penguin amplitudes.

\section{How accurate are the SU(3) relations? }
After analysing the rate of the SU(3) violation for different
elements of the $B\to PP$ amplitudes
we are now in a position to return to the relation
(\ref{rel1})
and calculate the magnitude of its violation
representing the individual amplitudes in terms of the factorizable parts
and nonfactorizable corrections. As we already mentioned,
in this particular
relation the penguin and annihilation contributions are absent.
We obtain:
\begin{equation}
\delta_{SU(3)}=\frac{(c_1+c_2/3+2c_2r_E^{B\pi K})f_K/f_\pi+
(c_2+c_1/3+2c_1r_E^{BK\pi})f_{BK}/f_{B\pi}}{
[(c_1+c_2/3+2c_2r_E^{B\pi\pi})+(c_2+c_1/3+2c_1r_E^{B\pi\pi})]}
-1\,.
\label{delta1}
\end{equation}

Using the numerical results for
$r_E^{B\pi\pi}$, $r_E^{B\pi K}$ and
$r_E^{B K \pi}$ obtained in the previous section
and the ratio of form factors (\ref{Bkbpi})
we obtain
\begin{equation}
\delta_{SU(3)}= (0.21^{+0.015}_{-0.014})
+\left(0.008_{-0.015}^{+0.013}\right)i
\label{deltares}.
\end{equation}
For consistency the Wilson coefficients $c_{1,2}$ have been
taken at the same scale $\mu_b$ at which the hadronic
matrix elements have been calculated from LCSR.
Importantly, our result for $\delta_{SU(3)}$ has a rather small
uncertainty indicating a moderate SU(3)-breaking in the relation (\ref{rel1})
which can be  taken into account in the applications of this relation.

To demonstrate that the situation is not always like that,
let us consider the U-spin relation
\begin{equation}
A(B_s\to K^+K^-) \simeq A(B_d\to\pi^+\pi^-)
\end{equation}
which is employed in certain CP-violation studies
\cite{FleischerPRep}.
From the results obtained above we are able to
predict the ratio
of factorizable hadronic matrix elements of $O_1$
for these channels (written without CKM factors).
\begin{equation}
\frac{A_{fact}(B_s\to K^+K^-)
}{A_{fact}(B_{d}\to
\pi^+\pi^-)}
=\left(\frac{f_K}{f_\pi}\right)
\left(\frac{f_{B_sK}(0)}{f_{B\pi}(0)}\right)\frac{m_{B_s}^2-m_K^2}{m_B^2-m_\pi^2}
= 1.76^{+0.15}_{-0.17}.
\label{Uspinrel}
\end{equation}
The nonfactorizable corrections to these relations
are more complicated and include annihilation and penguin
contributions which are not discussed here. We only notice
that the predicted violation of the U-spin is  quite substantial.
Note that on general grounds there is actually no preference for
U-spin symmmetry with respect to the  general SU(3).
Finally, with our results  one can also estimate the accuracy of the
other relation
\begin{equation}
A(B_s\to K^+K^-) \simeq A(B_d\to\pi^+K^-)
\end{equation}
suggested \cite{FleischerPRep} as an
estimate for the $B_s\to K^+ K^-$ amplitude.
We get (neglecting nonfactorizable corrections):
\begin{equation}
\frac{A(B_s\to K^+K^-)_{fact}}{A(B_d\to
\pi^+ K^-)_{fact}}
=\left(\frac{f_{B_sK}(0)}{f_{B\pi}(0)}\right)\frac{m_{B_s}^2-m_K^2}{m_B^2-m_\pi^2}
= 1.45^{+0.13}_{-0.14},
\label{Uspinrel1}
\end{equation}
again, a rather large SU(3)-violation effect.

\section{Conclusion}
We have demonstrated that
QCD sum rules provide {\em quantitative} estimates of 
$SU(3)$-violating corrections to the amplitude relations 
for charmless $B$ decays.
Our main goal was to formulate a consistent approach 
where all relevant hadronic matrix elements (decay constants, form factors
and hadronic decay amplitudes) are calculated 
with the same method (a combination of two-point and 
light-cone sum rules) and using a universal input ( quark masses,
condensates, meson distribition amplitudes).
The clear advantage of this approach is the possibility to 
calculate the flavour symmetry-violating corrections in terms of
the differences between the $s$ and $u,d$ quark masses and condensates.

For the SU(3) relation we have taken as a study case we predict
a moderate correction, with small uncertainties, indicating
that the method works, despite the fact that QCD sum rules have
limited accuracy.
Simultaneosly, we have demonstrated that, according to LCSR,
SU(3) violating effects in the heavy-light form factors
are not suppressed in the $m_b\to \infty $ limit.
Furthermore, the sum rule approach is able to identify the cases
where accumulation of several effects leads
to a large SU(3)-breaking, such as in the U-spin relation
between the factorizable amplitudes $B_s\to K^+K^-$
and $B \to \pi^+\pi^-$.
In such cases flavour symmetry is not reliable and an actual QCD calculation
for separate decay amplitudes is preferable.

The accuracy of our calculation
can still be improved, with a better knowledge of $m_s$
and the nonperturbative parameters of the kaon DA's
($a_1^K$ , $a_2^{K}$, $\delta_K^2$ etc.). Note that having at hand
precise measurements of $D\to K$ and kaon electromagnetic form factors
and comparing the sum rule predictions for these form factors with
the data, one may gain a lot of important constraints on these parameters
and improve the accuracy of the results obtained above.

\section*{Acknowledgments}

We are grateful to P.~Ball for communicating the results
of ref.\cite{BallB} prior to publication.
This work is supported by the DFG Sonderforschungsbereich
SFB/TR9 ``Computational Particle Physics'', and by the
German Ministry for Education and Research (BMBF).

\begin{appendix}
\section{ Normalization parameter of the twist 4 kaon DA }

The normalization parameter of the twist-four DA's of the
kaon has not been calculated yet. The corresponding
normalization for the pion  is given by the nonperturbative quantity
$\delta_\pi^2$, defined by
the matrix element
\begin{equation}
\bigl<0|\tilde{a}_\mu|\pi^+(p)\bigr>=-i f_\pi \delta_\pi^2 p_\mu
\end{equation}
of the current
\begin{equation}
\tilde{a}_\mu=\bar{d}\gamma_\rho\tilde{G}_{\rho\mu} u,
\end{equation}
where
$\tilde{G}_{\rho\mu}=\frac{1}{2}\epsilon_{\rho\mu\alpha\beta}G^{\alpha\beta}$.
It was determined in \cite{NSVVZ} using standard two-point sum rules. Two
different approaches were used, the results of which where shown to
be in a good agreement. The first one is based on the non-diagonal
correlator of $\tilde{a}_\mu$ with $j^{(\pi)}_\nu$
and is sensitive to the gluon condensate
density. We prefer to use the diagonal correlator
\begin{equation}
\tilde{\pi}_{\mu\nu}(q)=i\int
e^{iqx}d^4x\bigl<0|T\{\tilde{a}_\mu^\dag(x),\tilde{a}_\nu(0)\}|0\bigr>.
\end{equation}
In order to calculate $\delta_K^2$ one simply has to
replace $d\to s$ in the currents.
The correlator consists of two independent structures,
$\sim q_\mu q_\nu$ and
$\sim g_{\mu\nu}$, of which only the first one, denoted as
$\tilde{\pi}(q^2)$, is of interest.
Following the standard procedure with dispersion relation, quark-hadron
duality and Borel transformation, the sum rule is obtained:
\begin{equation}
\delta_K^4 f_K^2=\frac{1}{\pi}\int_0^{s_0^K}ds~\textrm{Im}_s\tilde{\pi}^{QCD}(s)
e^{(m_K^2-s)/M^2}.
\end{equation}
The intermediate hadronic states are the same as in the sum rule for
$f_K$, so that the hadronic threshold parameter $s_0^K$ and the Borel
window are fixed: $s_0^K= 1.05 ~\mbox{GeV}^2$, $0.5~\mbox{GeV}^2<M^2<1.2~\mbox{GeV}^2$.
For the calculation of the correlator in QCD, we take into account condensates
up to dimension 6, except the $d=5$ quark-gluon condensate which 
is suppressed. Also the perturbative part shown to be negligible in
\cite{NSVVZ} is left out. Our result reads
\begin{equation}
\delta_K^4 f_K^2=e^{m_K^2/M^2}\Bigg\{ M^2\Biggl[
\frac{\alpha_s m_s}{6\pi}
\left(\bigl<\bar{s}s\bigr>
-\frac{4}{3}  \bigl<\bar{u}u\bigr>
\right)
\end{equation}
\begin{equation}
+\frac{1}{72}\left<\frac{\alpha_s}{\pi}G^2\right>\Biggr]
+\frac{8}{9}\pi
\alpha_s\bigl<\bar{s}s\bigr>\bigl<\bar{u}u\bigr>
\Biggr\}
+O\left(m_s^2\right)+O\left(m_s\left<\bar{q}Gq\right>\right). \nonumber
\end{equation}
In the limit $m_s\to 0$, $m_K\to m_\pi\approx 0$ ,
the quark condensate
does not contribute  and this expression agrees with the
original result for $\delta_\pi^4 f_\pi^2$.

\end{appendix}

\end{document}